\documentclass{ifacconf}

\usepackage{graphicx}      
\usepackage{natbib}        
\usepackage{epstopdf}

\usepackage{amsfonts}
\usepackage{amsmath}
\usepackage{amssymb}
\usepackage[utf8]{inputenc} 
\usepackage{algorithm}
\usepackage{algpseudocode}
\usepackage{color}  

\renewcommand{\Re}{\mathbb{R}}

\usepackage{multirow}
\newcommand{\bmat}[1]{\begin{bmatrix}#1\end{bmatrix}}

\newtheorem{problem}[thm]{Problem}
\newtheorem{assumption}[thm]{Assumption}
\newtheorem{remark}[thm]{Remark}
\newtheorem{proposition}[thm]{Proposition}
\newtheorem{corollary}[thm]{Corollary}

\newcommand{\R}{\mathbb{R}}
\newcommand{\N}{\mathbb{N}}
\graphicspath{{figures/}}
\usepackage[algo2e,vlined,linesnumbered]{algorithm2e}

\begin{document}
\begin{frontmatter}
\title{Continuous and discrete abstractions for planning, applied to ship docking\thanksref{footnoteinfo}} 

\thanks[footnoteinfo]{
This work was supported by the Peder Sather Center for Advanced Study, a consortium of UC Berkeley and nine Norwegian academic institutions. It was also supported in part by the U.S. National Science Foundation grant ECCS-1906164, the U.S. Air Force Office of Scientific Research grant FA9550-18-1-0253, and Research Council of Norway through the Centres of Excellence funding scheme, project number 223254 AMOS, FRINATEK project 274441 UNLOCK, and MAROFF project 280655 ORCAS.}

\author[First]{Pierre-Jean Meyer}
\author[Second]{He Yin}
\author[Third]{Astrid H. Brodtkorb} 
\author[First]{Murat Arcak}
\author[Third]{Asgeir J. S{\o}rensen}

\address[First]{Department of Electrical Engineering and Computer Sciences, University of California, Berkeley, USA, {\tt\small $\{$pjmeyer, arcak$\}$@berkeley.edu}}
\address[Second]{Department of Mechanical Engineering, University of California, Berkeley, USA, {\tt\small he\_yin@berkeley.edu}}
\address[Third]{Centre for Autonomous Marine Operations (AMOS), Department of Marine Technology, Norwegian University of Science and Technology (NTNU), Otto Nielsens vei 10, 7052 Trondheim, Norway, {\tt\small $\{$astrid.h.brodtkorb, asgeir.sorensen$\}$@ntnu.no}}

\begin{abstract}                
We propose a hierarchical control framework for the synthesis of correct-by-construction controllers for nonlinear control-affine systems with respect to reach-avoid-stay specifications.
We first create a low-dimensional continuous abstraction of the system and use Sum-of-Squares (SOS) programming to obtain a low-level controller ensuring a bounded error between the two models.
We then create a discrete abstraction of the continuous abstraction and use formal methods to synthesize a controller satisfying the specifications shrunk by the obtained error bound.
Combining both controllers finally solves the main control problem on the initial system.
This two-step framework allows the discrete abstraction methods to deal with higher-dimensional systems which may be computationally expensive without the prior continuous abstraction.
The main novelty of the proposed SOS continuous abstraction is that it allows the error between abstract and concrete models to explicitly depend on the control input of the abstract model, which offers more freedom in the choice of the continuous abstraction model and provides lower error bounds than when only the states of both models are considered.
This approach is illustrated on the docking problem of a marine vessel.
\end{abstract}


\end{frontmatter}

\section{Introduction}
\label{sec introduction}
Abstraction-based control synthesis aims to abstract a system into a simpler model, synthesize a controller on the abstraction and finally refine this controller to ensure the satisfaction of the same control objective on the initial system.
Starting from a continuous initial system modeled as a differential equation, two abstraction-based control approaches can be considered.
In the \emph{hierarchical control} approach, we create a continuous abstraction with less variables or simpler dynamics than the initial model, and we create a low-level controller for the concrete model to track the abstract one~\citep{girard2009hierarchical}.
Note that this is slightly different from \emph{model reduction} in which the input and output variables of both models are kept identical~\citep{Antoulas_model_reduction_survey}.
In the \emph{symbolic control} approach, we create a discrete abstraction by partitioning the state space and using reachability analysis methods to over-approximate the continuous dynamics into a finite transition system~\citep[see e.g.][]{reissig2016feedback}.
Due to the state space partitioning, discrete abstractions are limited in their scalability.
One possible approach to reduce the complexity is to decompose the concrete system into smaller subsystems for which discrete abstractions are more easily created~\citep[see e.g.][]{pola2017decentralized}.
This method is applicable to weakly interconnected networked systems, but is not always practical for strongly interconnected systems with no clear structure to guide the decomposition.

In this paper, we address the scalability problem of discrete abstractions through an alternative approach, by considering a two-step process sketched in Figure~\ref{fig overview} and described in more details in Section~\ref{sub overview}.
In the first step, we design a continuous abstraction of the concrete model and use Sum-of-Squares (SOS) programming to find a low-level controller ensuring that the concrete model tracks trajectories of the continuous abstraction with an associated error bound.
Therefore, for the concrete system to satisfy a \emph{reach-avoid-stay} specification (reach a target set while avoiding unsafe sets, then stay there), it is sufficient to look for a controller of the continuous abstraction satisfying the same specification with sets shrunk by the error bound.
The second step aims to create a discrete abstraction of the lower-dimensional continuous abstraction and synthesize, using formal methods, a correct-by-construction controller to satisfy the shrunk specifications.

Although continuous and discrete abstractions are not novel ideas on their own, few results have attempted to combine them, and their applicability has been limited to restrictive classes of systems, such as a double integrator~\citep{fainekos2009temporal}, piecewise affine systems~\citep{mickelin2014synthesis}, differentially flat systems~\citep{colombo2013approximate} or bipedal robots~\citep{ames2015first}.
In contrast, the SOS-based continuous abstraction proposed here is applicable to the large class of control-affine nonlinear systems approximated with polynomial dynamics.
In addition to its broader applicability, the proposed method allows the error between abstract and concrete models to depend not only on the states of both models, but also on the control input of the abstract model.
This input dependence is particularly important when abstracting a dynamical model into its kinematic version, since we want to minimize the error between the velocities which are states of the concrete model and inputs of the abstract one.
More generally, this offers more freedom in the choice of the continuous abstraction model and provides lower error bounds than when only the abstract state is considered.

In comparison, existing continuous abstraction methods such as those relying on simulation functions for contracting systems~\citep{yang2014design}, Hamilton-Jacobi reachability analysis~\citep{herbert2017fastrack}, or SOS programming~\citep{Singh2018RobustTW,Stan2019} are all restricted to abstraction errors defined relative to the state variables and not the control inputs.
\cite{herbert2017fastrack} further combine their results with existing path planners similarly to our second step, including online methods such as RRT~\citep{kuffner2000rrt} which are computationally efficient but might not be appropriate for safety critical problems where satisfaction of the control objective needs to be guaranteed before taking any control action.
In contrast, we provide formal guarantees at the cost of increased computational complexity.

We apply this approach to a scenario where a marine vessel docks autonomously at a harbor. Today, this maneuver is done manually, due to high risk of collision and strict requirements for precision, even when system faults have occurred.
Typically, path planning for autonomous ships will consist of an offline algorithm making the preliminary plan based on available information like time and fuel consumption constraints, weather, and pre-defined safety margins, and an online part doing contingency-handling (e.g. collision avoidance).
In order for autonomous ships to be allowed to sail, the control system software must be verified so that it is at least as safe as human navigated ships \citep{DNVGL-AutoRemoteShips}.
By using correct-by-construction methods for design of offline path planning algorithms, the burden on simulation-based testing of the autonomous control system implementation is greatly reduced.

This paper is organized as follows.
Section~\ref{sec prelim} formulates the considered problem and provides an overview of the proposed two-step approach.
Section~\ref{sec continuous abstraction} presents the first step and main theoretical contribution of this paper on continuous abstraction.
Section~\ref{sec discrete abstraction} provides the discrete abstraction procedure of the second step, which is presented for self-containment of the overall approach.
Finally, the proposed method is illustrated in Section~\ref{sec ship} for the docking problem on the $6$-dimensional model of a marine vessel.

\section{Preliminaries}
\label{sec prelim}

\subsection{Notations}
\label{sub notations}
Let $\N$, $\R$ and $\R_+$ denote the sets of non-negative integers, real numbers and non-negative real numbers, respectively.
For $\xi \in \mathbb{R}^n$, $\mathbb{R}[\xi]$ represents the set of polynomials in $\xi$ with real coefficients, and $\R^{m}[\xi]$ and $\R^{m\times p}[\xi]$ denote all vector and
matrix valued polynomial functions.
The subset $\Sigma[\xi] = \{p = p_1^2 + p_2^2 + ... + p_M^2~|~p_1, ..., p_M \in \mathbb{R}[\xi]\}$ of $\mathbb{R}[\xi]$ is the set of SOS polynomials in $\xi$.
A set $X\subseteq\R^{n}$ is an interval of the vector space $\R^n$ if there exists $\underline x,\overline x\in X$ such that for all $x\in X$ we have $\underline x\leq x\leq \overline x$ using componentwise inequalities.
Given a positive vector $\varepsilon\in\R^n_+$ and a set $X\subseteq\R^{n}$, we introduce $X^{+\varepsilon}=\{x+e\in\R^n~|~x\in X,~e\in[-\varepsilon,\varepsilon]\}$ and $X^{-\varepsilon}=\{x\in\R^n~|~x+[-\varepsilon,\varepsilon]\subseteq X\}$ as the set $X$ expanded and shrunk by the interval $[-\varepsilon,\varepsilon]$, respectively.

\subsection{Problem formulation}
\label{sub pb}
Consider a control-affine nonlinear system
\begin{equation}
    \dot{x} = f(x, w) + g(x, w)u, 
    \label{eq system}
\end{equation}
with state $x \in X\subseteq\Re^{n_x}$, bounded control input $u \in U\subseteq \Re^{n_u}$, bounded disturbance input $w \in W\subseteq\Re^{n_w}$ and Lipschitz continuous functions $f: \Re^{n_x} \times \Re^{n_w} \rightarrow \Re^{n_x}$ and $g: \Re^{n_x} \times \Re^{n_w} \rightarrow \Re^{n_x\times n_u}$.
The sets $X$, $U$ and $W$ are assumed to be intervals of their respective spaces.

The control objectives are formulated as \emph{reach-avoid-stay} games which combine several safety and reachability sub-goals.
In addition to the state constraints defined by the set $X$, we define two subsets $X_a,X_r\subseteq X$, where the safety specification aims to avoid the set $X_a$ at all time, while the reach-stay objective is to reach the set $X_r$ in finite time and then stay there forever.

\begin{problem}
\label{pb reach avoid stay}
Given system (\ref{eq system}) and subsets $X_a,X_r\subseteq X$, find a set of initial states $X_0\subseteq X$ and a control strategy $u:X\rightarrow U$ such that for any disturbance signal $w:\R_+\rightarrow W$, all trajectories $x:\R_+\rightarrow\Re^{n_x}$ of the closed-loop system initialized in $X_0$ satisfies $x(t)\in X\backslash X_a$ for all $t\geq0$ and there exists $t_r\geq0$ such that $x(t)\in X_r$ for all $t\geq t_r$.
\end{problem}

Particular cases of safety, reachability, reach-avoid or reach-stay games can be considered by removing the corresponding conditions in Problem~\ref{pb reach avoid stay}.

\subsection{Overview of the proposed approach}
\label{sub overview}
\begin{figure}[tbh]
    \centering
    \includegraphics[width=\columnwidth]{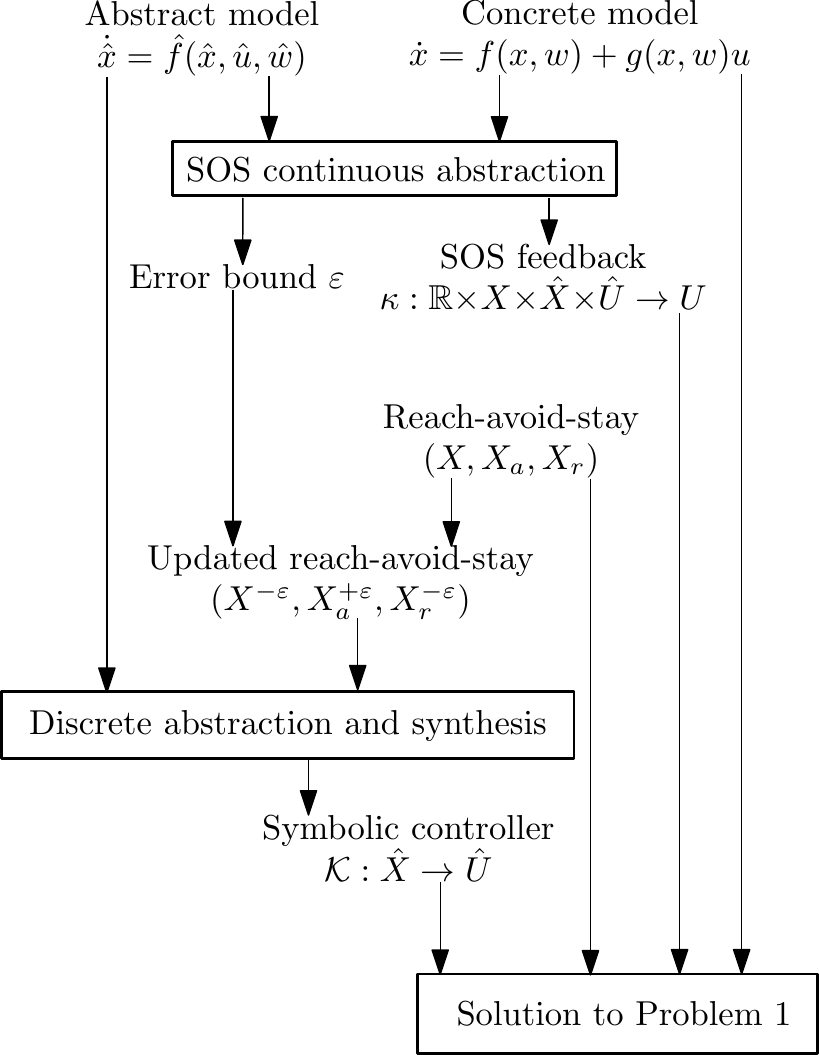}
    \caption{Overview of the design steps to solve Problem~\ref{pb reach avoid stay}.}
    \label{fig overview}
\end{figure}

In this paper, we solve Problem~\ref{pb reach avoid stay} in a two-step approach summarized below and in Figure~\ref{fig overview}, first by creating a continuous abstraction of the concrete model (\ref{eq system}), and next by using formal methods to synthesize a correct-by-construction controller on a discrete abstraction of the lower-dimensional continuous abstraction.

Given the concrete model (\ref{eq system}), the continuous abstraction
\begin{equation}
    \label{eq reduced}
    \dot{\hat x}=\hat f(\hat x,\hat u,\hat w),
\end{equation}
and the state and input constraints on both the concrete model ($X\subseteq\R^{n_x}$, $U\subseteq\R^{n_u}$, $W\subseteq\R^{n_w}$) and the abstract model ($\hat X\subseteq\R^{\hat n_x}$, $\hat U\subseteq\R^{\hat n_u}$, $\hat W\subseteq\R^{\hat n_w}$), the first step is to design a low-level controller $\kappa:\R\times X\times\hat X\times\hat U\rightarrow U$ ensuring that the concrete model (\ref{eq system}) can track trajectories of the abstract model (\ref{eq reduced}) with an error upper bounded by the known vector $\varepsilon\in\R^{n_x}$.
This is achieved by applying the Sum-of-Squares (SOS) methods detailed in Section~\ref{sec continuous abstraction}.

To solve the reach-avoid-stay specifications $(X,X_a,X_r)$ from Problem~\ref{pb reach avoid stay} on the concrete model (\ref{eq system}), it is thus sufficient to solve an auxiliary problem on the abstract model (\ref{eq reduced}) with respect to the reach-avoid-stay specification $(X^{-\varepsilon},X_a^{+\varepsilon},X_r^{-\varepsilon})$ with the shrunk state constraints $X^{-\varepsilon}$ and target set $X_r^{-\varepsilon}$ and the expanded set of states to be avoided $X_a^{+\varepsilon}$ (see Figure~\ref{fig simu} in Section~\ref{sec ship} for an illustration of these sets).
The second step of the approach, detailed in Section~\ref{sec discrete abstraction}, then consists in creating a discrete abstraction of the abstract model (\ref{eq reduced}) to synthesize a symbolic controller $\mathcal{K}:\hat X\rightarrow\hat U$ solving this auxiliary problem.

Combining the symbolic controller $\mathcal{K}:\hat X\rightarrow\hat U$ with the low-level controller $\kappa:\R\times X\times\hat X\times\hat U\rightarrow U$ then results in a controller solving Problem~\ref{pb reach avoid stay} on the concrete system.

\subsection{Considering more general specifications}
\label{sub LTL}
The control synthesis in the discrete abstraction step described in Section~\ref{sec discrete abstraction} is not the main focus of this paper since it relies on existing algorithms for finite transition systems.
In order to keep the description of this synthesis step as concise as possible, we thus restricted Problem~\ref{pb reach avoid stay} to \emph{reach-avoid-stay} specifications, for which simple fixed-point algorithms can be applied.
We note however that more general specifications, such as those expressed as Linear Temporal Logic formulas~\citep{baier2008principles}, can in principle be considered through the use of Rabin games~\citep{belta2017formal} which are computationally more expensive.

There also exists an alternative to consider more general specifications while avoiding the use of a Rabin game.
For this we should focus the first step on designing a continuous abstraction with simpler dynamics, and then replace the second step by discrete abstraction methods restricted to these simpler dynamics, such as single integrator models~\citep{kress2009temporal} or linear systems~\citep{kloetzer2008fully}, which result in \emph{deterministic} transition systems for which Linear Temporal Logic specifications are easier to handle.
This approach is not detailed further in this paper since we instead made the choice to present in Section~\ref{sec discrete abstraction} a discrete abstraction method applicable to general nonlinear dynamics, which allows for more freedom in the choice of the dynamics of the continuous abstraction.

\section{Continuous Abstraction}
\label{sec continuous abstraction}
The first step of the proposed approach is to create a simplified version of the concrete model (\ref{eq system}), referred to as \emph{continuous abstraction} or \emph{abstract model} and defined with hatted notations as in (\ref{eq reduced}), with state $\hat x\in\hat X\subseteq\R^{\hat n_x}$, control input $\hat u\in\hat U\subseteq\R^{\hat n_u}$ and disturbance $\hat w\in\hat W\subseteq\R^{\hat n_w}$.
Since the main goal of this first step is to reduce the complexity of the second step in Section~\ref{sec discrete abstraction} (exponential in the state-control dimension), we want to choose a continuous abstraction whose state and control dimensions satisfy $\hat n_x+\hat n_u < n_x+n_u$.

We introduce a map $\pi: \R^{\hat{n}_x}\times\R^{\hat{n}_u} \rightarrow \Re^{n_x}$ providing a reference trajectory to be followed by the concrete model, based on both the state and the control signals of the abstract model, while all other methods in the literature (see references in Section~\ref{sec introduction}) only rely on the abstract state.
We can then define the error $e \in \Re^{n_x}$ between trajectories of the concrete and abstract models:
\begin{align}
    e = x - \pi(\hat{x}, \hat{u}). \label{eq:ErrorState}
\end{align}

In this paper, we use affine maps $\pi(\hat{x}, \hat{u}) = P[\hat{x}; \hat{u}] + \Omega$, where matrix $P \in \Re^{n_x \times (\hat{n}_x + \hat{n}_u)}$ has at most one non-zero element per row, and $\Omega \in \Re^{n_x}$.

\begin{remark}
\label{rmk sos error}
Although this method can be used without any restriction on $P$, having one non-zero element per row is critical for the discrete abstraction approach in Section~\ref{sec discrete abstraction}. 
Indeed, the above restriction on the rows of $P$ is a sufficient condition to preserve intervals through the inverse image of $\pi$: i.e.\ if $X\subseteq\R^{n_x}$ is an interval of the concrete state space, then the set $\hat X\times\hat U=\{(\hat x,\hat u)\in\R^{\hat{n}_x}\times\R^{\hat{n}_u}~|~\pi(\hat x,\hat u)\in X\}$ is an interval in the abstract state-input space $\R^{\hat{n}_x}\times\R^{\hat{n}_u}$.
\end{remark}

The error dynamics resulting from (\ref{eq:ErrorState}) are given as
\begin{equation}
    \label{eq:error_dyn}
    \dot{e} = f_e(e, \hat{x}, \hat{u}, w, \hat{w}) + g_e(e, \hat{x}, \hat{u}, w) u - \frac{\partial \pi(\hat x,\hat u)}{\partial \hat{u}}\dot{\hat{u}},
\end{equation}
with $f_e(e,\hat{x},\hat{u},w,\hat{w}) = f(e+\pi(\hat{x},\hat{u}),w) - \frac{\partial \pi(\hat x,\hat u)}{\partial \hat{x}}\hat{f}(\hat{x}, \hat{u}, \hat{w})$ and $g_e(e,\hat{x},\hat{u}, w) = g(e+\pi(\hat{x},\hat{u}),w)$.
Let $E_0 \subseteq \Re^{n_x}$ denote a compact set of initial conditions for the error system (\ref{eq:error_dyn}).

In Section~\ref{sec discrete abstraction}, $\hat u$ is first designed as a discrete-time signal, then implemented in the abstract model (\ref{eq reduced}) with a zero-order hold. This means 
\begin{align}
    &\hat{u}(t) = \hat{u}(\tau_i), \ \forall t \in [\tau_i, \tau_{i+1}), \ \text{with} \ \tau_i = iT_s, \nonumber \\
    &\hat{u}(\tau_{i+1}) = \hat{u}(\tau_i) + \Delta\hat{u}(\tau_{i+1}), \label{eq:uhat_jump}
\end{align}
where $T_s$ is the sampling period, $\Delta\hat{u}(t)$ is the periodic change in the abstract control, restricted to a set $\Delta \hat{U} \subseteq \Re^{\hat{n}_u}$.
Since the signal $\hat{u}$ is piecewise constant, we thus have 
\begin{align}
\dot{\hat{u}}(t) = 0, \ \forall t \in [\tau_i, \tau_{i+1}). \nonumber 
\end{align}

We initially focus our analysis of the error system (\ref{eq:error_dyn}) on the first sampling period $[0, T_s)$, before the input jump $\Delta\hat u$ at time $T_s$.
Given the bounded set of initial conditions $E_0$, we want to enforce the boundedness of the error state during $[0, T_s)$ by introducing a low-level controller
\begin{align}
\label{eq sos controller}
u(t) = \kappa(t, e(t), \hat{x}(t), \hat{u}(t)),
\end{align}
defined by a time-varying, error-state feedback control law $\kappa: \Re \times \Re^{n_x} \times \Re^{\hat{n}_x} \times \Re^{\hat{n}_u} \rightarrow \Re^{n_u}$.
Below, we provide the design requirements on $\kappa$ to obtain such an error bound.
\begin{proposition}
\label{prop sos}
Given the error dynamics (\ref{eq:error_dyn}) 
and $\gamma \in \Re$, $T_s > 0$, $\hat{X} \subseteq \mathbb{R}^{\hat{n}_x}$, $\hat{U} \subseteq \mathbb{R}^{\hat{n}_u}$, $\hat{W} \subseteq \mathbb{R}^{\hat{n}_w}$, $W \subseteq \mathbb{R}^{n_w}$, if there exists a $\mathcal{C}^1$ function $V: \R \times \mathbb{R}^{n_x} \rightarrow \mathbb{R}$, and $\kappa: \R \times \mathbb{R}^{n_x} \times \mathbb{R}^{\hat{n}_x} \times \mathbb{R}^{\hat{n}_u} \rightarrow \mathbb{R}^{n_u}$, such that 
\begin{align}
    &E_0 \subseteq \{e~|~V(0,e)\leq \gamma\}, \label{eq:V_cond0}\\
    & \frac{\partial V(t,e)}{\partial e}(f_e(e, \hat{x}, \hat{u},w,\hat{w})+ g_e(e, \hat{x}, \hat{u},w)\kappa(t, e,\hat{x},\hat{u}))\nonumber \\
    &\ ~~~~ + \frac{\partial V(t,e)}{\partial t} \leq 0, \ \forall t, e, \hat{x}, \hat{u}, w, \hat{w}, \ \text{s.t.} \ t\in[0,T_s), \nonumber \\
	&\ \ ~~~~ V(t, e) = \gamma, \ \hat{x} \in \hat{X}, \ \hat{u} \in \hat{U},\ w \in W, \ \hat{w} \in \hat{W}, \label{eq:V_cond1}
\end{align}
then for all $e(0) \in E_0$, we have $e(t) \in \{e~|~V(t,e)\leq \gamma\}$, for all $t \in [0,T_s)$.
\end{proposition}

\begin{pf}
Let $V$ and $\kappa$ satisfy \eqref{eq:V_cond0}-\eqref{eq:V_cond1}.
Assume there exists $t_2\in[0,T_s)$, initial condition $e_0 \in E_0$ and signals $\hat{x}(t) \in \hat{X}$, $\hat{u}(t) \in \hat{U}$, $w(t) \in W$, $\hat{w}(t) \in \hat{W}$ such that a trajectory $e(t)$ from $e(0) = e_0$ satisfies $V(t_2,e(t_2))>\gamma$.
Since $V(0,e(0))\leq\gamma$ from \eqref{eq:V_cond0}, then by continuity of $V$ there exists $t_1\in[0,t_2)$ such that $V(t_1,e(t_1))=\gamma$ and $t_1=\inf_{\text{s.t. }V(t,e(t))>\gamma}t$, which contradicts $\dot{V}(t_1,e(t_1),\hat{x}(t_1),\hat{u}(t_1),w(t_1),\hat{w}(t_1))<0$ from \eqref{eq:V_cond1}.
\qed
\end{pf}

Although Proposition~\ref{prop sos} is stated for the first sampling period $[0,T_s)$, it can be used for any other sampling period $[\tau_i,\tau_{i+1})$ with $\tau_i=iT_s$. 

\begin{corollary} \label{coro:shift_V}
Define the funnel $F(t) = \{e~|~V(t,e)\leq \gamma\}\subseteq \Re^{n_x}$.
For all $e(\tau_{i}) \in F(0)$, we have $e(\tau_{i}+t) \in F(t)$, for all $t \in [0, T_s)$, under the control signal $u(\tau_i + t)=\kappa(t,e(\tau_i+t),\hat{x}(\tau_i+t),\hat{u}(\tau_i+t)))$.
\end{corollary}

Next, we focus on the effect of the input jump $\Delta\hat{u}$ at the end of each sampling period as in \eqref{eq:uhat_jump}.
From (\ref{eq:ErrorState}), $\Delta\hat{u}$ induces a jump on the error described as follows, where $\tau_{i+1}^-$ and $\tau_{i+1}^+$ denote sampling instant $\tau_{i+1}$ before and after the discrete jump, respectively:
\begin{align}
    e(\tau_{i+1}^+) =& \ x(\tau_{i+1}^+) - P[\hat{x}(\tau_{i+1}^+) ; \hat{u}(\tau_{i+1}^+)] - \Omega\nonumber \\
    =& \ x(\tau_{i+1}^-) - P[\hat{x}(\tau_{i+1}^-); \hat{u}(\tau_{i+1}^-)+\Delta \hat{u}(\tau_{i+1}^+)] - \Omega \nonumber \\
    =& \ e(\tau_{i+1}^-) - P\left[0; \Delta \hat{u}(\tau_{i+1}^+)\right]. \nonumber
\end{align}

We introduce the additional condition below to characterize the error jump induced by the control jump $\Delta\hat{u}$ in terms of the funnel from Corollary~\ref{coro:shift_V}.

\begin{proposition} \label{prop_jump}
Given $\gamma \in \Re$, $\Delta \hat{U} \in \Re^{\hat{n}_u}$, if there exists a function $V: \Re \times \Re^{n_x} \rightarrow \Re$ satisfying
\begin{align}
    &V(0, e - P[0;\Delta \hat{u}]) \leq \gamma, \ \forall e, \Delta\hat{u}, \nonumber \\
    &~~~~~~~~~~~~~~~~~~~~~~~~~~~~~ \ \text{s.t.} \ V(T_s,e) \leq \gamma, \ \Delta\hat{u} \in \Delta \hat{U}, \label{eq:V_cond2}
\end{align}
then for all $e(\tau_{i+1}^-) \in F(T_s)$, we have $e(\tau_{i+1}^+) \in F(0)$.
\end{proposition}

\begin{figure}[tbh]
    \centering
    \includegraphics[width=0.9\columnwidth]{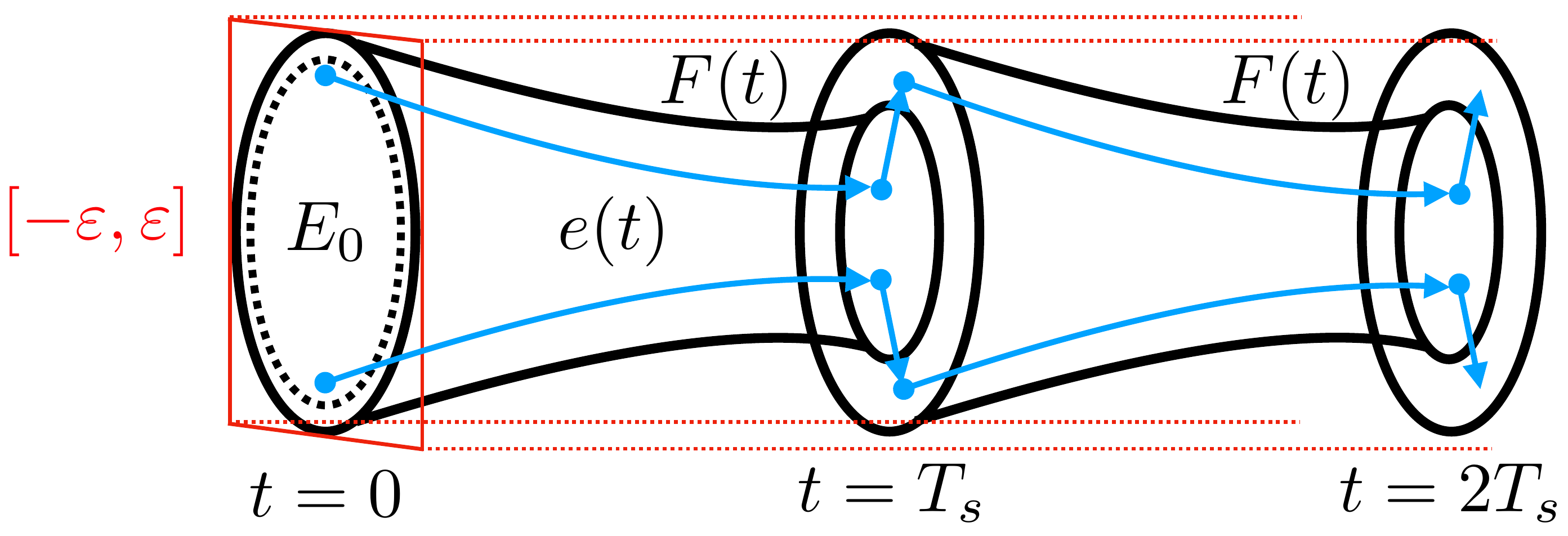}
    \caption{Illustration of Theorem \ref{th sos}, with initial error set $E_0$, funnels $F$ on each sampling period, bounded error jumps at sampling times. The red interval hull $[-\varepsilon,\varepsilon]$ of $\cup_{t\in[0,T_s)}F(t)$ bounds the error $e(t)$ for all times.}
    \label{fig: funnel}
\end{figure}

We next combine the conditions for the error boundedness for each sampling period and discrete jump from Propositions~\ref{prop sos} and~\ref{prop_jump}, respectively, to obtain the main result on the boundedness of the error at all time, formulated below and illustrated in Figure~\ref{fig: funnel}.

\begin{thm}
\label{th sos}
If there exists $V$ and $\kappa$ satisfying \eqref{eq:V_cond0}--\eqref{eq:V_cond2}, define $\varepsilon\in\R_+^{n_x}$ such that $\cup_{t\in[0,T_s)}F(t)\subseteq[-\varepsilon,\varepsilon]$.
Then for all $t \ge 0$, $\hat{x}(t) \in \hat{X}$, $\hat{u}(t) \in \hat{U}$, $\Delta\hat{u}(t) \in \Delta \hat{U}$, $w(t) \in W$ and $\hat{w}(t) \in \hat{W}$, the error system (\ref{eq:error_dyn}) under control law $u(t)=\kappa(\tilde{t},e(t),\hat{x}(t),\hat{u}(t)))$ with $\tilde t=(t\mod T_s)\in[0,T_s)$ satisfies:
$$e(0) \in E_0\Rightarrow\forall t \ge 0,~e(t)\in[-\varepsilon,\varepsilon].$$
\end{thm}
\begin{pf}
From Corollary \ref{coro:shift_V} and for all $\tau_i=iT_s$, we have if $e(\tau_i) \in F(0)$, then $e(\tau_i+\tilde{t}) \in F(\tilde{t})$ and $e(\tau_{i+1}^-) \in F(T_s)$. Then it follows from Proposition \ref{prop_jump} that $e(\tau_{i+1}^+) \in F(0)$. As a result, for all $e(0) \in E_0 \subseteq F(0)$, we have $e(kT_s +\tilde{t}) \in F(\tilde{t}) \subseteq [-\varepsilon,\varepsilon]$, for all $k \ge 0$, and $\tilde{t} \in [0, T_s)$. \qed
\end{pf}

Finding storage functions $V$ and control laws $\kappa$ satisfying the constraints \eqref{eq:V_cond0}--\eqref{eq:V_cond2} is a difficult problem.
In this paper, we use SOS programming to search for them at the cost of a restriction to polynomial candidates $V \in \Re[(t,x)]$ and $\kappa \in \Re^{n_u}[(t,x,\hat{x},\hat{u})]$. We make the following assumptions besides the requirement that system \eqref{eq system} is control-affine.
\begin{assumption}
The error dynamics \eqref{eq:error_dyn} are polynomials: $f_e \in \Re^{n_x}[e,\hat{x},\hat{u},w,\hat w]$ and $g_e \in \Re^{n_x\times n_u}[e,\hat{x},\hat{u},w]$.
$E_0$, $\hat{X}$, $\hat{U}$, $\Delta \hat{U}$, $W$ and $\hat W$ are semi-algebraic sets: there exists $p_0 \in \Re[e]$ such that $E_0 = \{e \in \Re^{n_x}~|~p_0(e) \leq 0\}$; with similar definitions for $\hat{X}$, $\hat{U}$, $\Delta \hat{U}$, $W$, $\hat W$ with polynomials $p_{\hat x} \in \Re[\hat{x}]$, $p_{\hat u} \in \Re[\hat{u}]$, $p_\Delta \in \Re[\Delta\hat{u}]$, $p_w \in \Re[w]$, $p_{\hat{w}} \in \Re[\hat{w}]$.
\end{assumption}

For a general nonlinear system, least-squares regression, Taylor expansion and change of variables can be used to obtain a polynomial system~\citep[see e.g.][]{Antonis:02}. By applying the generalized S-procedure \citep{Parrilo:00} to \eqref{eq:V_cond0}--\eqref{eq:V_cond2}, and choosing the volume of $F(t)$ as the cost function, we obtain the following optimization problem:
\begingroup
\allowdisplaybreaks
\begin{subequations} \label{eq:sosopt}
\begin{align}
\displaystyle \min_{V, \kappa, s,l} ~&
        \int_0^{T_s} \text{volume}(F(t))dt \vspace{2mm} \nonumber \\ 
        \ \  \mathrm{s.t.} ~
    &s_i \in \Sigma[(t,e,\hat{x},\hat{u},w,\hat{w})], \ \forall i \in \{1,...,4\}, 
    \nonumber \\
    &s_j \in \Sigma[(e,\Delta\hat{u})], \ \forall j \in \{5,6\},\nonumber \\
    & l \in \Re[(t,e,\hat{x},\hat{u},w,\hat{w})], s_0 \in \Sigma[e], \label{eq:sos_cond0}\\
    &-(V(0,e) - \gamma) + s_0\cdot p_0 \in \Sigma[e], \label{eq:sos_cond1}\\
    &-\left(\frac{\partial V}{\partial t}+\frac{\partial V}{\partial e}\cdot(f_e+g_e \kappa)\right) +l \cdot (V - \gamma) + s_1 \cdot p_{\hat x} \nonumber \\
    &~~~~~~~~ +s_2 \cdot p_{\hat u} +s_3 \cdot p_w  - s_4 \cdot t (T_s - t) \nonumber \\
    &~~~~~~~~ \in \Sigma[(t,e,\hat{x},\hat{u},w,\hat{w})], \label{eq:sos_cond2} \\
    &-(V(0,e - P[0;\Delta \hat{u}])-\gamma) + s_5\cdot(V(T_s,e)-\gamma) \nonumber \\
    &~~~~~~~~ +s_6 \cdot p_{\Delta} \in \Sigma[(e,\Delta\hat{u})], \label{eq:sos_cond3}
\end{align}
\end{subequations}
\endgroup
where $s_i$, $s_j$ and $l$ are S-procedure certificates.

The above optimization has three bilinear pairs of decision variables: $(\frac{\partial V}{\partial e}, \kappa)$, $(l, V)$, $(s_5, V(T_s,e))$, making it non-convex. If we either fix $V$, or $(\kappa, l, s_5)$, the constraints in \eqref{eq:sosopt} are convex. Similar to \cite{Stan2019}, we tackle the optimization above by alternating the search over $V$ and $(\kappa, l, s_5)$, and solving a series of convex problems as shown in Algorithm \ref{alg:alg1}.
In the $\gamma$-step, the volume of $F(t)$ is shrunk by minimizing $\gamma$; in \eqref{eq:funnel_contain} of the $V$-step, the funnel certified by the $V$-step, $\{e~|~V^j(e,t) \leq \gamma^j\}$, is enforced to be contained by the funnel from the $\gamma$-step, $\{e~|~V^{j-1}(e,t) \leq \gamma^j\}$,  for all $t \in [0, T_s]$.
For more details and an algorithm for the initialization of $V^0$ in Algorithm \ref{alg:alg1}, the reader is referred to~\cite{Stan2019}.

\begin{algorithm2e}[tbh]
	\KwData{Function $V^0$ such that  \eqref{eq:sos_cond1}--\eqref{eq:sos_cond3} are feasible by proper choice of $s, \kappa, l$. Tolerance $\delta_{\text{tol}} > 0$.}
	\KwResult{($\kappa$, $\gamma$, $V$) sub-optimal solution of (\ref{eq:sosopt}).}

	\Repeat{$\vert \gamma^{j} - \gamma^{j-1} \vert \leq \delta_{\textnormal{tol}}$}
	{
	$\boldsymbol{\gamma}$\textbf{-step}: decision variables $(s, l, \kappa,\gamma)$.
			
	 Minimize $\gamma$ subject to \eqref{eq:sos_cond0}--\eqref{eq:sos_cond3} using $V = V^{j-1}$. 
	 
	 This yields ($s_5^j, l^j, \kappa^j$) and optimal cost $\gamma^j$.
	 
	 $\boldsymbol{V}\textbf{-step}$: decision variables $(s_1\text{--}s_4, s_{6}, V)$.
			
	 Maximize the feasibility subject to \eqref{eq:sos_cond1}--\eqref{eq:sos_cond3} as 
	 
	 well as $s_a, s_b \in \Sigma[(t,x)],$ and
		\begin{align}
		& -s_a \cdot (V^{j-1} - \gamma^j) + (V - \gamma^j) &\nonumber \\
		&~~~~~~~~~~~~~~~~~~~~~~~~~~ - s_b\cdot t(T_s - t)\in \Sigma[(t,x)], \label{eq:funnel_contain}
		\end{align}
		
		 using ($\gamma = \gamma^j, s_5 = s_5^j, l = l^j, \kappa = \kappa^j$). This yields $V^j$.
	
    }
    \caption{Iterative optimization of (\ref{eq:sosopt}) to obtain function $V$ and low-level controller $\kappa$ satisfying (\ref{eq:V_cond0})-(\ref{eq:V_cond2}).}
	\label{alg:alg1}
\end{algorithm2e}

After finding the funnel $F(t)$ characterized by $V$ from Algorithm~\ref{alg:alg1}, computing the interval error bound $[-\varepsilon, \varepsilon]\subseteq\R^{n_x}$ is achieved by solving the optimization problem: $\min_{\varepsilon} \sum_{i = 1}^{n_x} \varepsilon_i$, s.t. $F(t) \subseteq [-\varepsilon, \varepsilon]$ for all $t \in [0, T_s)$, which can be formulated as a convex SOS problem.
Once $\varepsilon$ is known, Theorem~\ref{th sos} implies that Problem~\ref{pb reach avoid stay} on the concrete model (\ref{eq system}) with the reach-avoid-stay specification $(X,X_a,X_r)$ can be solved through an auxiliary problem on the abstract model (\ref{eq reduced}) with respect to the modified reach-avoid-stay specification $(X^{-\varepsilon},X_a^{+\varepsilon},X_r^{-\varepsilon})$ using the notations from Section~\ref{sub notations} to shrink the state constraints $X^{-\varepsilon}$ and target set $X_r^{-\varepsilon}$ and expand the set of states to be avoided $X_a^{+\varepsilon}$ (see Figure~\ref{fig simu} in Section~\ref{sec ship} for an illustration of these sets).
Since $X$, $X_a$, $X_r$ are intervals of $\R^{n_x}$, the updated sets are also intervals.
We can then define $\hat X^\varepsilon,\hat X_a^\varepsilon,\hat X_r^\varepsilon\subseteq\R^{\hat n_x}$ and $\hat U^\varepsilon,\hat U_a^\varepsilon,\hat U_r^\varepsilon\subseteq\R^{\hat n_u}$ as the projections of these sets into the abstract state-input space $\R^{\hat n_x}\times\R^{\hat n_u}$ using the inverse image of $\pi:\R^{\hat n_x}\times\R^{\hat n_u}\rightarrow\R^{n_x}$,
\begin{align*}
\hat X^\varepsilon\times\hat U^\varepsilon &= \{(\hat x,\hat u)\in\R^{\hat n_x}\times\R^{\hat n_u}~|~\pi(\hat x,\hat u)\in X^{-\varepsilon}\},\\
\hat X_a^\varepsilon\times\hat U_a^\varepsilon &= \{(\hat x,\hat u)\in\R^{\hat n_x}\times\R^{\hat n_u}~|~\pi(\hat x,\hat u)\in X_a^{+\varepsilon}\},\\
\hat X_r^\varepsilon\times\hat U_r^\varepsilon &= \{(\hat x,\hat u)\in\R^{\hat n_x}\times\R^{\hat n_u}~|~\pi(\hat x,\hat u)\in X_r^{-\varepsilon}\}.
\end{align*}
As mentioned in Remark~\ref{rmk sos error}, due to the restriction of the affine map $\pi$ in (\ref{eq:ErrorState}), all these hatted sets are intervals.

\begin{problem}
\label{pb reduced}
Given system (\ref{eq reduced}) and subsets $\hat X_a^\varepsilon,\hat X_r^\varepsilon\subseteq\hat X^\varepsilon$ and $\hat U_a^\varepsilon,\hat U_r^\varepsilon\subseteq\hat U^\varepsilon$, find a set of initial states $\hat X_0\subseteq \hat X^\varepsilon$ and a control strategy $\hat\kappa:\hat X^\varepsilon\rightarrow \hat U^\varepsilon\backslash\hat U_a^\varepsilon$ such that for any disturbance signal $\hat w:\R_+\rightarrow\hat W$, all trajectories $\hat x:\R_+\rightarrow\Re^{\hat n_x}$ of the closed-loop system initialized in $\hat X_0$ satisfies $\hat x(t)\in \hat X^\varepsilon\backslash \hat X_a^\varepsilon$ for all $t\geq0$ and there exists $\hat t_r\geq0$ such that $\hat x(t)\in \hat X_r^\varepsilon$ and $\hat\kappa(\hat x(t))\in\hat U_r^\varepsilon$ for all $t\geq \hat t_r$.
\end{problem}

\section{Discrete abstraction and synthesis}
\label{sec discrete abstraction}
In Section~\ref{sec continuous abstraction}, we defined a continuous abstraction (\ref{eq reduced}) and obtained from Theorem~\ref{th sos} a bound $\varepsilon\in\R^{n_x}$ on the error with respect to the concrete model (\ref{eq system}).
In this section, the second step of the proposed approach is to solve Problem~\ref{pb reduced} by creating a discrete abstraction of (\ref{eq reduced}) and using classical model checking tools to synthesize a satisfying controller.

The discrete abstraction of (\ref{eq reduced}) takes the form of a finite transition system defined as a triple $(\mathcal{X},\mathcal{U},\delta)$ containing a finite set of states $\mathcal{X}$, a finite set of inputs $\mathcal{U}$, and a non-deterministic transition relation $\delta:\mathcal{X}\times\mathcal{U}\rightarrow2^{\mathcal{X}}$.
We first take a finite partition of $\hat X^\varepsilon\backslash\hat X_a^\varepsilon$ into smaller intervals, where each partition element is represented by a unique discrete state (also called \emph{symbol}) in $\mathcal{X}$.
We add a last symbol $Out\in\mathcal X$ corresponding to the complement set $Out=\R^{\hat n_x}\backslash(\hat X^\varepsilon\backslash\hat X_a^\varepsilon)$, so that $\mathcal X$ becomes a partition of the whole state space $\R^{\hat n_x}$.
Although uniform grid-like partitions are most commonly used, arbitrary partitions into intervals are also admissible.
No formal result currently exists in the literature on how to choose the granularity of the partition, but some empirical guidelines are provided in Section~\ref{sec ship}.
Next, we define $\mathcal{U}$ as a finite discretization of the set of admissible control inputs $\hat U^\varepsilon\backslash\hat U_a^\varepsilon$.

Before defining the transition relation $\delta$, we first introduce $\hat x(t;\hat x_0,\hat u,\hat w)$ to denote the state reached at time $t\geq0$ by the abstract model (\ref{eq reduced}) starting in $\hat x_0$, with constant control input $\hat u$ and with disturbance signal $\hat w:[0,t]\rightarrow\hat W$.
Then for an interval of initial states $[a,b]\subseteq\R^{\hat n_x}$, the finite time reachable set of (\ref{eq reduced}) is defined as
\begin{equation*}
    \hat R(t,[a,b],\hat u)=\{\hat x(t;\hat x_0,\hat u,\hat w)~|~\hat x_0\in[a,b],~\hat w:[0,t]\rightarrow\hat W\}.
\end{equation*}
Then, given the time sampling $T_s$ from Proposition~\ref{prop sos}, the set of successors associated to each pair $(s,\hat u)\in(\mathcal{X}\backslash\{Out\})\times\mathcal{U}$ is defined as
\begin{equation}
    \label{eq transition}
    \delta(s,\hat u)=\{s'\in\mathcal{X}~|~s'\cap\hat R(T_s,s,\hat u)\neq\emptyset\}.
\end{equation}
Since the true reachable set $\hat R(T_s,s,\hat u)$ can rarely be computed exactly, we can replace it in (\ref{eq transition}) by an over-approximation.
Using over-approximations preserves the fact that any behavior of (\ref{eq reduced}) can be reproduced by the discrete abstraction, which in turns ensures that a controller synthesized on the discrete abstraction also satisfies the desired specifications on (\ref{eq reduced}).
Several methods for the over-approximation of reachable sets using intervals and applicable to most nonlinear systems are described in~\cite{meyer2019hscc}.
Problem~\ref{pb reduced} on (\ref{eq reduced}) is then translated into a control problem on its discrete abstraction.

\begin{problem}
\label{pb abstraction}
Find a set of initial symbols $\mathcal{X}_0\subseteq\mathcal{X}\backslash\{Out\}$ and a control strategy $\mathcal{K}:\mathcal{X}\rightarrow \mathcal{U}$ such that any closed-loop trajectory of the discrete abstraction $s_0,s_1,\dots$ (i.e.\ such that $s_0\in\mathcal{X}_0$ and $s_{i+1}\in\delta(s_i,\mathcal{K}(s_i))$ for all $i\geq1$) satisfies $s_i\in\mathcal{X}\backslash\{Out\}$ for all $i\geq0$, and there exists $k\in\N$ such that $s_i\subseteq\hat X_r^\varepsilon$ and $\mathcal{K}(s_i)\in\hat U_r^\varepsilon$ for all $i\geq k$.
\end{problem}

The control synthesis on the discrete abstraction is achieved in two stages: first a safety game in the target set for the \emph{stay} part of the specifications, then a reachability game with respect to the resulting safe set for the \emph{reach-avoid} part.
Both these games are solved through classical fixed-point algorithms, outlined below and in Algorithm~\ref{algo abstraction}.
These algorithms are known to terminate in finite time and reach the maximal fixed-points~\citep{tabuada2009verification}.

For the safety game (lines 1-5 in Algorithm~\ref{algo abstraction}), we compute a controlled invariant subset of the set of symbols fully included in the target interval $\{s\in\mathcal{X}~|~s\subseteq\hat X_r^\varepsilon\}$.
The loop is initialized with this target set and iteratively removes symbols which cannot be kept in the current set $S$.
The \emph{stay} specification on the control input is achieved by restricting this loop to inputs in $\mathcal{U}\cap\hat U_r^\varepsilon$.
The loop stops once it reaches a fixed-point, and we can define the \emph{stay} controller $\mathcal{K}:S\rightarrow\mathcal{U}\cap\hat U_r^\varepsilon$ by using any valid control input from the last iteration of the loop.

Next, the reachability game (lines 6-10) is initialized with the set $S\subseteq\{s\in\mathcal{X}~|~s\subseteq\hat X_r^\varepsilon\}$ resulting from the safety game and iteratively expands it with all the symbols that can be forced to reach the current set $R$ in a single step.
The loop stops when a fixed-point is reached, and the \emph{reach-avoid} part of the controller $\mathcal{K}:R\backslash S\rightarrow\mathcal{U}$ is defined by using any valid input from the last iteration of the loop.
Note that the \emph{avoid} parts of the specification are automatically satisfied by defining $\mathcal{U}\subseteq\hat U^\varepsilon\backslash\hat U_a^\varepsilon$ and ensuring in Algorithm~\ref{algo abstraction} that the safety and reachability sets cannot contain the symbol $Out$: $S\subseteq R\subseteq\mathcal{X}\backslash\{Out\}$.

\begin{algorithm2e}[tbh]
  \KwData{Discrete abstraction $(\mathcal{X},\mathcal{U},\delta)$, target sets $\hat X_r^\varepsilon$, $\hat U_r^\varepsilon$.}
  {\bf Safety game initialization:} $S\leftarrow \{s\in\mathcal{X}~|~s\subseteq\hat X_r^\varepsilon\}$\\
  \Repeat{$S$ reaches a fixed-point}{$S\leftarrow \{s\in S~|~\exists\hat u\in\mathcal{U}\cap\hat U_r^\varepsilon,~\delta(s,\hat u)\subseteq S\}$}
  Extract $\mathcal{K}:S\rightarrow\mathcal{U}\cap\hat U_r^\varepsilon$ satisfying the last loop iteration.\\
  {\bf Reachability game initialization:} $R\leftarrow S$\\
  \Repeat{$R$ reaches a fixed-point}{$R\leftarrow \{s\in \mathcal{X}\backslash\{Out\}~|~\exists\hat u\in\mathcal{U},~\delta(s,\hat u)\subseteq R\}$}
  Extract $\mathcal{K}:R\backslash S\rightarrow\mathcal{U}$ satisfying the last loop iteration.
\caption{Control synthesis for a \emph{reach-avoid-stay} game on the discrete abstraction $(\mathcal{X},\mathcal{U},\delta)$.\label{algo abstraction}}
\end{algorithm2e}

The final step of the overall approach is to refine the controllers $\mathcal{K}:\mathcal{X}\rightarrow\mathcal{U}$ from Algorithm~\ref{algo abstraction} and $\kappa: \R \times \mathbb{R}^{n_x} \times \mathbb{R}^{\hat n_x} \times \mathbb{R}^{\hat n_u} \rightarrow \mathbb{R}^{n_u}$ from Section~\ref{sec continuous abstraction} into a controller solving Problem~\ref{pb reach avoid stay} for the concrete system (\ref{eq system}).
We first introduce the function $H:\R^{\hat n_x}\rightarrow\mathcal{X}$ such that $H(\hat x)=s\Leftrightarrow\hat x\in s$, mapping each state of the continuous abstraction (\ref{eq reduced}) to the unique symbol of the discrete abstraction containing it.
We can then define a controller $\hat \kappa:\R\times\hat X^\varepsilon\rightarrow\hat U^\varepsilon\backslash\hat U_a^\varepsilon$ for (\ref{eq reduced}) as the zero-order hold version of $\mathcal{K}:\mathcal{X}\rightarrow\mathcal{U}$ with sampling period $T_s$. For all index $k\in\N$ and time $t\in\R_+$ such that $kT_s\leq t<(k+1)T_s$, we have
\begin{equation}
    \label{eq controller reduced}
    \hat\kappa(t,\hat x(t))=\mathcal{K}(H(\hat x(kT_s))).
\end{equation}
Combining (\ref{eq controller reduced}) with the low-level controller $\kappa: \R \times \mathbb{R}^{n_x} \times \mathbb{R}^{\hat n_x} \times \mathbb{R}^{\hat n_u} \rightarrow \mathbb{R}^{n_u}$ from (\ref{eq sos controller}), we obtain a controller $C:\R\times\R^{n_x}\times\R^{\hat n_x}\rightarrow\R^{n_u}$ for the concrete model defined as
\begin{equation}
    \label{eq controller initial}
    C(t,x,\hat x)=\kappa(t,x-\pi(\hat x,\hat\kappa(t,\hat x)),\hat x,\hat\kappa(t,\hat x)).
\end{equation}
Since this controller also depends on the abstract state $\hat x(t)$, its use in the concrete model (\ref{eq system}) requires the computation of a trajectory of the closed-loop continuous abstraction, as stated in the main result of this paper below.

\begin{thm}
\label{thm controller}
Given $E_0\subseteq\R^{n_x}$ bounding the initial error state (which is a design parameter in Proposition~\ref{prop sos}), the set of winning initial states for Problem~\ref{pb reach avoid stay} is
$X_0=\{\pi(\hat x,\mathcal{K}(H(\hat x)))\in\R^{n_x}~|~H(\hat x)\in R\}+E_0$.
Given an initial state $x_0\in X_0$, let $\hat x:\R\rightarrow\R^{\hat n_x}$ be any trajectory of the continuous abstraction (\ref{eq reduced}) with controller $\hat\kappa$ in (\ref{eq controller reduced}) initialized in $\{\hat x_0\in\bigcup_{s\in R}s~|~x_0-\pi(\hat x_0,\mathcal{K}(H(\hat x_0)))\in E_0\}$.
Then, the closed-loop system (\ref{eq system}) with controller (\ref{eq controller initial}) satisfies the reach-avoid-stay specification from Problem~\ref{pb reach avoid stay}.
\end{thm}
\begin{pf}
By construction in Algorithm~\ref{algo abstraction}, the controller $\mathcal{K}:\mathcal{X}\rightarrow\mathcal{U}$ solves Problem~\ref{pb abstraction} for the discrete abstraction $(\mathcal{X},\mathcal{U},\delta)$ with a winning set of initial states $R\subseteq\mathcal{X}\backslash\{Out\}$.

From the definition of the transition relation $\delta$ in (\ref{eq transition}), it can be shown as in~\cite{tabuada2009verification} that the function $H:\R^{\hat n_x}\rightarrow\mathcal{X}$ is an alternating simulation relation between the discrete abstraction $(\mathcal{X},\mathcal{U},\delta)$ and the continuous abstraction (\ref{eq reduced}).
This implies that if the discrete abstraction is controlled with $\mathcal{K}:\mathcal{X}\rightarrow\mathcal{U}$ from Algorithm~\ref{algo abstraction}, then the zero-order hold version of this controller (\ref{eq controller reduced}) gives behavior of the continuous abstraction that can all be reproduced by the discrete abstraction.
Since $\mathcal{K}$ solves Problem~\ref{pb abstraction} for the discrete abstraction, we can deduce that the trajectories $\hat x$ of the closed-loop continuous abstraction remain at all time within the set $\bigcup_{s\in R}s\subseteq\hat X^\varepsilon\backslash\hat X_a^\varepsilon$ with controls $\hat\kappa(t,\hat x(t))\in\mathcal{U}\subseteq\hat U^\varepsilon\backslash\hat U_a^\varepsilon$.
In addition, there exists $k\in\N$ such that for all $t\geq kT_s$ we have $\hat x(t)\in\bigcup_{s\in S}s\subseteq\hat X_r^\varepsilon$ and $\hat\kappa(t,\hat x(t))\in\hat U_r^\varepsilon$.
Therefore, $\hat\kappa$ solves Problem~\ref{pb reduced} with the winning set of initial states $\hat X_0=\bigcup_{s\in R}s$.

If $x_0\in X_0$ as defined in the theorem statement, then there exists a winning state of the continuous abstraction $\hat x_0\in\bigcup_{s\in R}s$ such that $x_0-\pi(\hat x_0,\mathcal{K}(H(\hat x_0)))\in E_0$.
From Theorem~\ref{th sos}, we thus know that for any trajectories $x$ of (\ref{eq system}) controlled with (\ref{eq controller initial}) and $\hat x$ of (\ref{eq reduced}) controlled with (\ref{eq controller reduced}), we have $e(t)=x(t)-\pi(\hat x(t),\hat\kappa(t,\hat x(t)))\in[-\varepsilon,\varepsilon]$ for all $t\geq 0$.
Since the trajectories $(\hat x(t),\hat\kappa(t,\hat x(t)))$ satisfy the reach-avoid-stay specification in Problem~\ref{pb reduced} defined by the sets $(X^{-\varepsilon},X_a^{+\varepsilon},X_r^{-\varepsilon})$, we can deduce that the closed-loop trajectory $x(t)$ of system (\ref{eq system}) satisfies the initial reach-avoid-stay specification $(X,X_a,X_r)$ from Problem~\ref{pb reach avoid stay}.
\qed
\end{pf}

\section{Case study: marine vessel}
\label{sec ship}
\begin{figure}[tbh]
\centering
\includegraphics[width=\columnwidth]{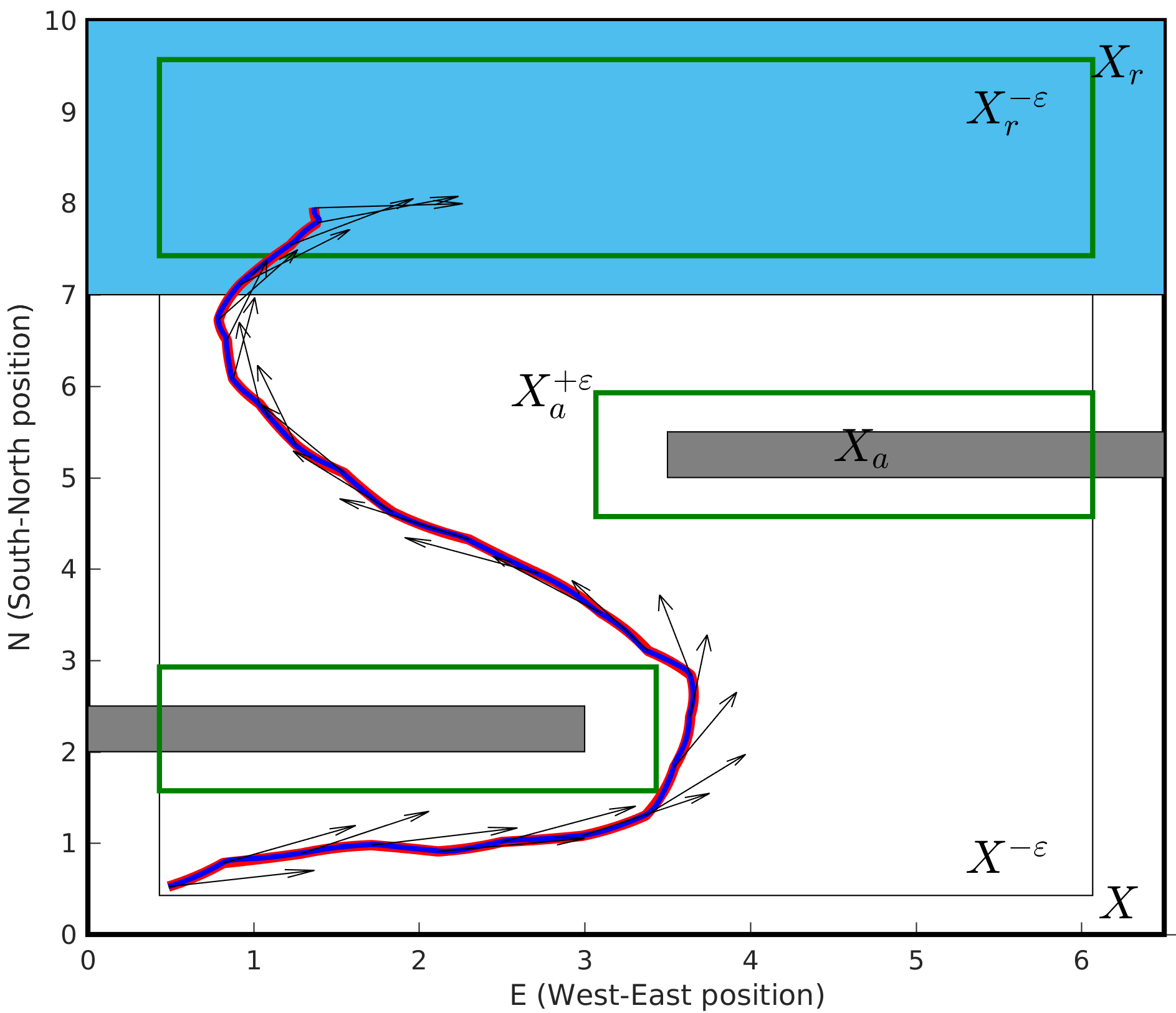}
\caption{Closed-loop trajectories of the abstract (red) and concrete (blue) models in the $(N,E)$-plane with the ship heading $\psi$ (black arrows), the initial and shrunk state constraints $X$ and $X^{-\varepsilon}$ (thick and thin black lines), the target set $X_r$ (light blue), the obstacles $X_a$ (grey) and the shrunk target set $X_r^{-\varepsilon}$ and expanded obstacles $X_a^{+\varepsilon}$ (green).
\label{fig simu}}
\end{figure}

The autonomous docking maneuver consists of four phases: transit, transition from high speed to low speed maneuvering, docking, and dockside keeping a steady contact force with the dock.
In this work we focus on the transition phase, which is challenging due to large changes in the ship dynamics when the speed is reduced.
This means that unlike the general Problem~\ref{pb reach avoid stay}, we only consider a \emph{reach-avoid} specification to reach the area near the dock (light blue in Figure~\ref{fig simu}) while avoiding the piers (gray areas).
The \emph{stay} part of the specification is omitted as it is handled in the later docking and dockside phases.

The ship motion at moderate speed can be modeled as in~\cite{Fossen}:
\begin{subequations} \label{eq:initial_shipmodel}
\begin{align}
\dot{\eta} &= R(\psi) \nu + v_c, \label{eq:fhat}\\
M \dot{\nu} + C(\nu) \nu + D \nu &= \tau + R(\psi)^\top \tau_{wind},
\end{align}
\end{subequations}
where $\eta = [N;E;\psi]$ are the South-North and West-East positions and heading of the ship ($\psi=0$ points North, $\psi=\pi/2$ points East), $\nu =[u;v;r]$ are the surge and sway velocities, and yaw rate of the ship.
$R(\psi) = \left[ \begin{smallmatrix}
\cos(\psi) & -\sin(\psi) & 0 \\
\sin(\psi) & \cos(\psi) & 0 \\
0 & 0 & 1
\end{smallmatrix} \right]$ is a rotation matrix.
$\tau\in\R^3$ is the control input affecting the three acceleration states of the ship.
$v_c\in\R^3$ and $\tau_{wind}\in\R^3$ are disturbances corresponding to current velocities and wind forces.
The inertia matrix including hydrodynamic added mass $M=\left[ \begin{smallmatrix}
87.4 & 0 & 0\\
0 & 98.3 & 2.48\\
0 & 2.48 & 22.2
\end{smallmatrix} \right]$, damping matrix $D=\left[ \begin{smallmatrix}
6.58 & 0 & 0\\
0 & 37.7 & 2.66\\
0 & 2.66 & 19.3
\end{smallmatrix} \right]$ and Coriolis matrix $C(\nu) = \nu(1) \left[ \begin{smallmatrix} 0&0&0\\0&0&98.3\\0&0&2.48\end{smallmatrix} \right]$ are chosen for a $1:30$ scale model of a platform supply vessel.

Using the notations from (\ref{eq system}), we have state $x=[\eta;\nu]\in\R^6$, control input $u=\tau\in\R^3$ and disturbance input $w=[v_c;\tau_{wind}]\in\R^6$.
The controls are unconstrained ($U=\R^3$) and the disturbances signals are assumed to be evolve in $W=[-0.01,0.01]^5\times[-0.05,0.05]$.
The chosen reach-avoid specification focuses on the first three states with the safety constraints $X=[0,10]\times[0,6.5]\times[-\pi,\pi]\times\R^3$, the obstacles $X_a=X_{a1}\cup X_{a2}$ with $X_{a1}=[2,2.5]\times[0,3]\times[-\pi,\pi]\times\R^3$ and $X_{a2}=[5,5.5]\times[3.5,6.5]\times[-\pi,\pi]\times\R^3$ (in grey in Figure~\ref{fig simu}), and the target set $X_r=[7,10]\times[0,6.5]\times[\pi/3,2\pi/3]\times\R^3$ (light blue).

The continuous abstraction is chosen as the kinematics part of the concrete model \eqref{eq:initial_shipmodel}:
\begin{align}
    \label{eq ship reduced}
    \dot{\hat{\eta}} &= R(\hat{\psi}) \hat{\nu} + \hat{v}_c
\end{align}
where the abstract states, inputs and disturbances are $\hat{x} = \hat{\eta}$, $\hat{u} = \hat{\nu}$ and $\hat{w} = \hat{v}_c$.
The map $\pi$ is chosen as $\pi(\hat{x},\hat{u}) = [\hat{x}; \hat{u}]$.
However, instead of defining error as in \eqref{eq:ErrorState}, we redefine the error state as $e = \phi\cdot (x - \pi(\hat{x},\hat{u}))$, where $\phi = \bmat{R^{-1}(\hat{\psi}), \boldsymbol{0}_{3\times 3}; \boldsymbol{0}_{3\times 3}, \boldsymbol{I}_{3}}$.
The matrix $\phi$ allows to replace the trigonometric functions in $\hat{\psi}$ in the error dynamics \eqref{eq:error_dyn} by trigonometric functions in $e(3) = (\psi - \hat{\psi})$, which can easily be approximated by polynomials in certain range of $e(3)$.
The input, input jump, and disturbances spaces for the abstract model are $\hat{U} = [0, 0.18] \times [-0.05, 0.05] \times [-0.1, 0.1]$, $\Delta\hat{U} = [-0.18, 0.18] \times [-0.1, 0.1] \times [-0.2, 0.2]$, and $\hat{W} = [-0.01, 0.01]^3$.
Algorithm~\ref{alg:alg1} is run with degree-2 polynomials to characterize the storage function $V$, control law $\kappa$, and multipliers $s, l$, and terminates in 6 minutes on a computer with 3.6GHz processor and 62GB of RAM.
The resulting error bounds $\varepsilon$ on $(N, E, \psi)$ are $[-0.427, 0.427] \times [-0.432, 0.432] \times [-0.235, 0.235]$ and the expanded obstacles $X_a^{+\varepsilon}$ and shrunk target set $X_r^{-\varepsilon}$ are outlined in green in Figure~\ref{fig simu}.
Due to the consideration of the abstract control $\hat u=\hat\nu$ in the error definition (\ref{eq:ErrorState}), the obtained error bounds are less conservative than those computed using \cite{Singh2018RobustTW}, that is $[-0.462, 0.462] \times [-0.493, 0.493] \times [-0.339, 0.339]$.

For the discrete abstraction as in Section~\ref{sec discrete abstraction}, we take a uniform partition of $\hat X$ into $50$ intervals per dimension (resulting in $|\mathcal{X}|=125000$) and a uniform discretization of $\hat U$ into $9$ values per dimension (i.e.\ $|\mathcal{U}|=729$).
To define the transition relation $\delta$ as in (\ref{eq transition}), we compute interval over-approximations of the reachable set of the continuous abstraction (\ref{eq ship reduced}) using the continuous-time mixed-monotonicity approach implemented in the tool TIRA~\citep{meyer2019hscc}.
The choice of the partition granularity with respect to the sampling period $T_s=3$ was done so that the reachable set would jump on average two to three partition cells away from its initial cell.
This ensures that the transitions do not jump too far, while also avoiding self-loops which hinder the synthesis.
On a server with $24$ cores at $2.5$GHz and $128$GB of RAM, the abstraction is created in $10$ seconds and the control synthesis is achieved after $15$ hours, resulting in a winning set $R\subseteq\mathcal{X}$ covering $93\%$ of the set of symbols $\mathcal{X}$.
Although these computation times may appear to be large, we emphasize that our whole approach is done offline with respect to static obstacles.
In addition, we highlight the significant complexity reduction of the continuous abstraction prior to the discrete abstraction by applying the approach in Section~\ref{sec discrete abstraction} directly to the full model (\ref{eq:initial_shipmodel}), which took over a week of computation on the same server with a coarse partition of $19$ intervals per dimension and resulting in a winning set coverage of only $0.04\%$ of $\mathcal{X}$.

The synthesized controller is then converted into the controllers (\ref{eq controller reduced}) and (\ref{eq controller initial}) for the abstract (\ref{eq ship reduced}) and concrete ship models (\ref{eq:initial_shipmodel}), respectively.
The initial state is chosen as a random point in the bottom left corner of the $(N,E)$-plane, and both closed-loop trajectories with random disturbance signals are plotted in red for (\ref{eq ship reduced}) and blue for (\ref{eq:initial_shipmodel}) in Figure~\ref{fig simu}.
The black arrows represent the orientation $\psi$ of the ship at each discrete time step.
We can first note that the low-level controller (\ref{eq sos controller}) provides a very efficient tracking of the abstract model's trajectory (red) by the concrete model (blue).
Both models satisfy their reach-avoid specifications by reaching the (shrunk) target set in blue while avoiding the (expanded) obstacles in grey.
Once the ship has reached the desired $[N;E]$ position (blue set) but not the correct orientation $\psi$, we can see it slowly drift sideways while it turns to face East.

\section{Conclusion}
\label{sec conclusion}
In this paper, we proposed a hierarchical framework combining continuous and discrete abstraction methods for the synthesis of correct-by-construction controllers for nonlinear control-affine systems with respect to reach-avoid-stay specifications.
In the first step, we create a low-dimensional continuous abstraction of a system and use Sum-of-Squares programming to obtain a low-level controller enforcing a maximum error bound between the two models.
The main novelty of this contribution is that the abstraction error is defined based on not only the states of both models, but also the control input of the abstract model, which offers more freedom in the choice of the continuous abstraction model and provides lower error bounds.
The second step then creates a discrete abstraction of the continuous abstraction (at a lower computational cost than if done on the initial model) and uses formal methods to synthesize a controller satisfying the specifications shrunk by the error bound.
Combined with the low-level controller, we finally obtain a controller satisfying the main specifications on the initial model.
This approach is illustrated on the docking problem of a marine vessel whose dynamics have too many state variables for the discrete abstraction approach to be applied directly.
The next step of this work is an experimental validation of the ship docking results in the $40\times 6.5$ m basin of the Marine Cybernetics Laboratory at NTNU.

\bibliographystyle{ifacconf}
\bibliography{Bibliography}

\end{document}